\documentclass[aps,showpacs,amsmath,amssymb]{revtex4}
\usepackage{graphicx}

\usepackage{graphicx}

\usepackage{dcolumn}
\usepackage{epsfig}
\usepackage{verbatim}

\def\be{\begin{equation}}
\def\ee{\end{equation}}
\newcommand{\ba}{\begin{array}}
\newcommand{\ea}{\end{array}}
\newcommand{\bea}{\begin{eqnarray}}
\newcommand{\eea}{\end{eqnarray}}

\newcommand{\op}[1]{\ensuremath{\widehat{ #1} }}
\newcommand{\der}{\partial}
\newcommand{\vct}[1]{\ensuremath\mbox{\boldmath$ #1 $}}
\newcommand{\mat}[1]{\ensuremath\mbox{$ \mathbb #1 $}}

\newcommand{\HH}{ \mathcal H}
\newcommand{\GO}{ \mathcal O}
\newcommand{\N}{ \mathcal N}
\newcommand{\vx}{\vct x}
\newcommand{\vy}{\vct y}

\newcommand{\vxi}{\vct \xi}
\newcommand{\veta}{\vct \eta}

\newcommand{\vl}{\vct l}

\newcommand{\mM}{\mat M}

\begin{document}

\title{Markovian evolution of Gaussian states in the semiclassical limit } 

\author{O.~Brodier$^{1}$ and A.~M.~Ozorio de Almeida$^{2}$}
\affiliation{$^1$ Laboratoire de Math\'ematiques et Physique Th\'eorique,
Facult\'e des Sciences et Techniques, Universit\'e de Tours, 37200 TOURS \\
$^2$ Centro Brasileiro de Pesquisas F\'isicas, 
Rua Dr. Xavier Sigaud, 150, 22290-180 Rio de Janeiro, BRASIL}

\date{\today}

\begin{abstract}

We derive an approximate Gaussian solution of the Lindblad equation in the semiclassical limit, given a general Hamiltonian and linear coupling with the environment. The theory is carried out in the chord representation and describes the evolved quantum characteristic function, which gives direct access to the Wigner function and the position representation of the density operator by Fourier transforms. The propagation is based on a system of non-linear equations taking place in a double phase space, which coincides with Heller's theory of unitary evolution of Gaussian wave packets when the Lindbladian part is zero. 

\end{abstract}


\maketitle

\section{Introduction}

Open quantum systems deal with physical systems which interact with their environment. From this interaction the system may loose or gain energy; this is the dissipation phenomenon. Coupling of a system with its environment entails loss of information in the sense that, if one starts with a pure state $\op\rho(0)$, such that its {\it purity}, 
$\mathop{Tr}\op\rho(0)^2 = 1$, then it will undergo a non-unitary evolution which will not preserve its purity, that is,  $\mathop{Tr}\op\rho(t)^2 < 1$. One important aspect of this loss is {\it decoherence}, that is the vanishing of the off-diagonal terms of the density operator. 

The Markovian process offers a powerfull mathematical tool in the study of this class of systems. It relies on the assumption that the environmental degrees of freedom are very fast as compared to the proper dynamics of the system, so that the future evolution of the density operator is a function of its value in the present only, regardless of its past. It was shown by a series of works, concluded by Lindblad in \cite{Lin76}, that the corresponding evolution equation can always be written in the form 
\begin{equation}
\frac{\der \op\rho}{\der t} = -\frac{i}{\hbar}\Big[ \op H, \op\rho \Big]
+\frac{1}{2 \hbar}\sum_k\left(
2 \op L_k\op\rho\op L_k^\dagger - L_k^\dagger\op L_k\op\rho - 
\op\rho L_k^\dagger \op L_k
\right).
\end{equation}
The Hamiltonian $\op H$ describes the unitary evolution of the system without environment, and Lindblad's theory proves the a priori existence of the operators $\op L_k$, which are commonly dubbed Lindblad operator, and which modelize the effect of the environment. The master equation of quantum optics is a particular, well understood case, where the Lindblad operators are $\op L_1 = \op a$, i.e. the annihilation operator, describing the emission process, and $\op L_2 = \op a^\dagger$ the absorption. One can see that these Lindblad operators are not Hermitian, which can be generalized to every case where the coupling to the environment is dissipative.

In this paper, we derive the consistent dynamics of a Gaussian solution of the Lindblad equation. As compared to former paper where we generalized the analysis of the evolution of extended states, that goes back to Van Vleck \cite{Van Vleck} (see also \cite{Gutzwiller}\cite{Maslov}\cite{Almeida}), this work is rather a generalization of the evolution of ``wave packets'', developed by Heller \cite{Heller} and Littlejohn \cite{Littlej}, among others. The approximation holds as long as the size of the Gaussian is small enough to justify an identification of the Hamiltonian with its second order expansion, that is, its quadratic kernel. Since the effect of decoherence is to quickly decrease the extension of the solution, the regime of validity should be larger than in the unitary case.

We chose to represent the Gaussian solution in the chord reresentation, that is, the Fourier transform of the Weyl-Wigner representation. After giving the form of the Lindblad equation in the chord frame, we assume a Gaussian expression of the solution, with time dependent parameters, and its insertion in the Lindblad equation leads to a system of equations of motion for these parameters. Comparison with Heller's theory is then derived in a slightly simplified case.

This paper presents a simpler, albeit less exact theory than \cite{BroAlm09}, which generalizes complex WKB approximations for the solutions of the Lindblad equation. In that treatment it is necessary to further complexify the doubled phase space. Thus, it is likely that the simpler approximations developed here may be of more practical use.

The theory can be useful for any initial state, which can be conceived as a finite decomposition into Gaussian ones.

All the following formulae are written for a system with a single degree of freedom, in order to clarify the notation. Nonetheless, it is quite simple to generalize them for a finite number of degrees of freedom.

\section{Equation in the Weyl formalism}

The Weyl representation maps every quantum operator onto a phase space function, that is, a function of the vector $\vx=(p,q)$\cite{Wey31}\cite{Alm98}. For an operator $\op A$ the Weyl symbol $A$ is defined as
\begin{equation}
A(\vx)= 2 \int \exp{(-\frac{i}{\hbar}p Q )}
~ \langle q +\frac{Q}{2}| \op A | q - \frac{Q}{2}\rangle~d Q.
\end{equation}
The Weyl symbol of the state operator $\op\rho$ is the Wigner function
\begin{equation}
W(\vx) = \N\int \exp{(-\frac{i}{\hbar}pQ )}
~\langle q+\frac{Q}{2}|\op \rho|q-\frac{Q}{2}\rangle ~dQ,
\end{equation}
with $\N = 1/(2\pi\hbar)$, and its Fourier transform, the chord function $\chi(\vct \xi)$, also called characteristic function, is
\begin{equation}
\chi(\vct \xi) = \N \int 
\exp{(-\frac{i}{\hbar}\vct \xi\wedge \vct x )}
~W(\vct x)~d\vct x,
\label{fourierwigner}
\end{equation}
where the wedge product of two vectors $\vct x=(p,q)$ and $\vct x'=(p',q')$ is defined by $\vct x\wedge \vct x'=pq'-p'q = \mat J \vct x \cdot \vct x'$, which also defines the skew matrix $\mat J = \left(\begin{array}{cc} 0 & -1 \cr 1 & 0 \end{array}\right)$. One can have the chord symbol directly from the quantum operator through the formula
\begin{equation}
\chi(\vct \xi) = \N \int 
\exp{(-\frac{i}{\hbar}\xi_p q)}
~\langle q+\frac{\xi_q}{2}|\op \rho|q-\frac{\xi_q}{2}\rangle ~dq.
\label{chordfunc}
\end{equation}
We call the space of all $\vx$ the centre space, and the space of all
$\vct\xi$ the chord space. 

In the chord space, by using product rules for the product of operators\cite{Alm98}, the Lindblad equation is represented by a partial differential equation. This equation is actually simpler than in the Weyl (centre) representation, and this justifies our choice. In the case where the Lindblad operators are linear functions of $\op p$ and $\op q$, that is, $\op L = \vct l'\cdot\op{\vct x} + i\vct l''\cdot\op{\vct x}$ with $\vct l'$ and $\vct l''$ real vectors, this equation can be written
\begin{eqnarray}
\frac{\der \chi}{\der t} (\vct \xi,t) & = & -\frac{i}{\hbar}
\N \int \Bigl[H(\vx'+\frac{1}{2}\vct \xi,t)
-H(\vx'-\frac{1}{2}\vct \xi,t)
\Bigr]
~\exp{ \left(\frac{i}{\hbar}(\vxi'-\vxi)\wedge \vx' \right)}
~\chi(\vct \xi',t)
~d\vct \xi'~d\vx' \cr
~ & ~ & -\gamma ~\vct \xi\cdot \frac{\der \chi}{\der \vct\xi}(\vct \xi,t) 
- \frac{1}{2\hbar}~\Bigl[
(\vl'\cdot\vct \xi)^2 + (\vl''\cdot\vct \xi)^2 \Bigr] 
~\chi(\vct \xi,t). 
\label{dynchi}
\end{eqnarray}
The {\it dissipation coefficient}, 
\begin{equation}
\gamma = \vl''\wedge\vl',
\end{equation}
is null for a Hermitian Lindblad operator ($\vl''=\vct 0$) and we then have a purely diffusive case. $H$ is the Weyl representation of the Hamiltonian of the isolated system and coincides with the corresponding classical Hamiltonian, up to corrections coming from non-commutativity of $\op p$ and $\op q$. Its arguments in equation (\ref{dynchi}) are the pair of remarkable points $\vct x_+ = \vct x + \frac{\vct \xi}{2}$ and $\vct x_- = \vct x - \frac{\vct \xi}{2}$, which can be considered as both tips of a chord $\vct \xi$. Although this chord $\vct \xi = (\vct\xi_p,\vct\xi_q)$ can be interpreted as an auxiliary conjugate variable of $\vct x$, in the current approach it is actually more convenient to write the solution in terms of $\vct y = \mat J\vct\xi = (-\vct\xi_q,\vct\xi_p)$. The direct sum of these conjugate spaces can indeed be interpreted as a double phase space, where $\vct x$ formally plays the role of the position $q$, and $\vy$ the role of its Fourier conjugate, the momentum $p$. Then the above equation becomes
\begin{eqnarray}
\frac{\der \chi}{\der t} (\vct y,t) & = & -\frac{i}{\hbar} 
\N \int \Bigl[H(\vx'- \frac{1}{2}\mat J\vct y ,t)
-H(\vx' + \frac{1}{2}\mat J\vct y,t)
\Bigr] 
~\exp{ \left(\frac{i}{\hbar}(\vct y'- \vct y)\cdot \vx' \right)}
~\chi(\vct y',t)
~d\vct y'~d\vct x' \cr
~ & ~ &
-\gamma ~\vct y \cdot \frac{\der \chi}{\der \vct y}(\vct y,t) 
- \frac{1}{2\hbar}
(\vct \lambda\cdot\vct y)^2~\chi(\vct y,t). 
\label{dynchi_y}
\end{eqnarray}
The same name has been kept for the characteristic function $\chi(\vct y,t)$, though strictly this should be $\chi(\vct \xi,t)=\chi(-\mat J \vct y,t)$, and we have set $\vct\lambda =
{\mat J} (\vct l' + i\vct l'')$. The term $\vct y \cdot \frac{\der \chi}{\der \vct y}$ can actually be included in the integral term, thanks to an integration by parts of the exponential, and one has finally
\begin{equation}
\frac{\der \chi}{\der t} (\vct y,t) = -\frac{i}{\hbar} 
\N \int \HH (\vct x',\vct y,t)
~\exp{ \left(\frac{i}{\hbar}(\vct y'- \vct y)\cdot \vx' \right)} 
~\chi_t(\vct y')
~d\vct y'~d\vct x' 
- \frac{1}{2\hbar}(\vct \lambda\cdot\vct y)^2 
~\chi(\vct y,t), 
\label{dynchi_synth}
\end{equation}
with
\begin{eqnarray}
\HH(\vct x',\vct y,t) & = & H(\vct x' - \frac{1}{2}\mat J \vct y,t)
-H(\vct x' + \frac{1}{2}\mat J \vct y,t) 
- \gamma ~ \vct x'\cdot\vct y \cr
~ & = & \HH^+(\vct x',\vct y,t) - \HH^-(\vct x',\vct y,t) - \gamma ~ \vct x'\cdot\vct y.
\label{defHH}
\end{eqnarray}
This is exactly the double Hamiltonian that generates the classical motion underlying the semiclassical approximations in \cite{AlmBro07}. Obviously, the double hamiltonian will be time-independent if it is obtained from a time-independent single Hamiltonian. Furthermore, in the absence of dissipation, both 
$\HH^+(\vct x,\vct y)=H(\vct x - \frac{1}{2}\mat J \vct y)$ and
$\HH^-(\vct x,\vct y)=H(\vct x + \frac{1}{2}\mat J \vct y)$ 
will also be constants which generate independent motions for both chord tips.

Notice that (\ref{dynchi_synth}) can also be written as
\begin{equation}
\frac{\der \chi}{\der t} (\vct y,t) = -\frac{i}{\hbar} 
\N \int \HH (\vct x',\vct y,t) 
~\exp{ (-\frac{i}{\hbar}\vct y\cdot \vx' )}
~W_t(\vct x')
~d\vct x' 
- \frac{1}{2\hbar}(\vct \lambda\cdot\vct y)^2 
~\chi(\vct y,t), 
\label{dynchi_synth_W}
\end{equation}
in terms of the evolving Wigner function, or, alternatively, as
\begin{equation}
\frac{\der \chi}{\der t} (\vct y,t) = 
-\frac{i}{\hbar}
\HH (-\frac{\hbar}{i}\frac{\der}{\der \vct y}^{(1)},\vct y^{(2)},t)
~\chi(\vct y,t)
- \frac{1}{2\hbar}(\vct \lambda\cdot\vct y)^2 
~\chi(\vct y,t), 
\label{dynchi_der_H}
\end{equation}
where $^{(1)}$ and $^{(2)}$ mean that the derivatives are taken first and then the $\vct y$ terms are multiplied.

The differential term in the RHS of (\ref{dynchi_der_H}) 
(or the integral term in the RHS of (\ref{dynchi_synth_W}))  
represents the unitary part of the evolution. In other words,
in the chord representation, the commutator is
\begin{equation}
\Bigl[\op H,\op \rho\Bigr]_{\textrm chord}
=\HH (-\frac{\hbar}{i}\frac{\der}{\der \vct y}^{(1)},
\vct y^{(2)}, t) \;\chi(\vct y,t).
\label{unitary}
\end{equation}

\section{General Dynamics}

We here assume that the chord representation $\chi(\vct y,t)$ of the localized state has the usual semiclassical form 
\begin{equation}
\chi(\vct y,t) = \exp{\frac{i}{\hbar}S(\vct y,t)},
\end{equation}
where $S(\vct y,t)$ is a function with complex values of order $\GO(\hbar^0)$.
It is important to notice that, if we find the time evolution of such a state determined by the Lindblad equation, then we can evolve any linear combination of such states, which can be coherent states for instance. This is a consequence of the linearity of the Lindblad equation.

This semiclassical form naturally induces an $\hbar$ expansion of the unitary part of the equation, as it is shown in the appendix \ref{appa},
\begin{equation}
\HH (-\frac{\hbar}{i}\frac{\der}{\der \vct y}^{(1)},
 \vct y^{(2)}, t) ~\chi(\vct y,t) = 
\Biggl[
\HH \left(-\frac{\der S}{\der \vct y}(\vct y,t), \vct y, t\right) + \GO(\hbar)
\Biggr] ~\chi(\vct y,t).
\end{equation}
Hence, by expanding (\ref{dynchi_der_H}) at leading order in $\hbar$, one obtains
the Hamilton Jacobi equation
\begin{equation}
\frac{\der S}{\der t}(\vct y,t)  = 
-\HH\left( -\frac{\der S}{\der \vct y}(\vct y,t), \vct y ,t\right)  
+ \frac{i}{2}( \vct \lambda\cdot\vct y )^2 + \GO(\hbar).
\label{eq_ordre1_complex}
\end{equation}
This is a double phase space generalization of the complex WKB theory in \cite{Hub03}.
In this paper, we will not develop a complex resolution of this equation, presented in \cite{BroAlm09}. We rather separate $S(\vct y,t) = A(\vct y,t) + iB(\vct y,t)$ into its real and imaginary parts, and get
\begin{equation}
\frac{\der A}{\der t}(\vct y,t) +i \frac{\der B}{\der t}(\vct y,t) = 
-\HH\left( -\frac{\der A}{\der \vct y}(\vct y,t) - i\frac{\der B}{\der \vct y}(\vct y,t),\vct y ,t\right)  
+ \frac{i}{2}( \vct \lambda\cdot\vct y )^2 + \GO(\hbar).
\label{eq_ordre1}
\end{equation}

\section{Gaussian characteristic function: consistent evolution}

\label{consistentgauss}

An initial Gaussian state will keep its Gaussian form if it evolves according to a quadratic Hamiltonian dynamics. That means that one must expand the Hamiltonian of (\ref{eq_ordre1}) up to order $2$ in $\vct x$ and $\vct y$ to obtain a consistent Gaussian evolution. Since these variables correspond to the coordinates of the trajectories supporting the state, the expansion will be faithful as long as the state is localized in a sufficiently small region of phase space. A coherent or even a squeezed state will obviously fulfill this condition in the semiclassical limit. This is the basic idea behind the following treatment, which can be seen as a double phase space generalization of Heller's theory of Gaussian wave packet evolution\cite{Heller}.

We assume that the chord function has the form
\begin{equation}
\chi(\vct y,t) = {\mathcal K} 
~\exp{\left(
\frac{i}{\hbar}a_t - \frac{i}{\hbar}(\vct y-\vct Y_t)\cdot \vct X_t
- \frac{1}{\hbar} b_t 
-\frac{1}{2\hbar}(\vct y-\vct Y_t)\cdot(\mM_t-i\mat N_t)(\vct y-\vct Y_t)\right)},
\label{chigauss}
\end{equation}
where $\mM_t$ and $\mat N_t$ are symmetric matrices. We have therefore, in the notation of the previous section,
\begin{eqnarray}
A(\vct y,t) & = & a_t - (\vct y-\vct Y_t)\cdot \vct X_t + \frac{1}{2} (\vct y-\vct Y_t)\cdot\mat N_t (\vct y-\vct Y_t)\cr
B(\vct y,t) & = & b_t + \frac{1}{2} (\vct y-\vct Y_t)\cdot\mM_t (\vct y-\vct Y_t),
\label{A_B_quad}
\end{eqnarray}
where $\vct Y_t$ represents the minimum of $B$, or the maximum of the modulus of the wave packet.

It is instructive to compare the above expressions with the position representation of the familiar wave packets, corresponding to (linearly) squeezed and rotated coherent states in \cite{Heller,Littlej}. Considering the analogy
between the underlying double phase space coordinates $(\vct y,\vct x)$ with the familiar phase space variables $(q, p)$,
we identify the matrix $\mM_t$ as describing the overall squeezing, i.e. its eigenvalues describe the compression in the chord space, $\vct y$, (compensated by stretching in $\vct x$), or vice versa. On the other hand, the matrix $\mat N_t$
accounts for the rotation in double phase space that tilts this 4-dimensional gaussian.  

Since $\frac{\der B}{\der \vct y}(\vct Y_t,t)=0$, it is natural to expand the complex equations (\ref{eq_ordre1}) around $\vct Y_t$, which leads to a separation of its real and imaginary parts. By identifying the terms $(\vct y - \vct Y_t)^n$ with $n=0,1,2$, one obtains
\begin{eqnarray}
\dot{\vct Y}_t & = & -\frac{\der\HH}{\der\vct x} - (\mM_t)^{-1}  \mat D \vct Y_t \cr
\dot{\vct X}_t & = &  \frac{\der \HH}{\der \vct y}  + \mat N_t (\mat M_t)^{-1} \mat D \vct Y_t \cr
\dot{\mat N}_t & = &  - \mat N_t \frac{\der^2 \HH}{\der \vct x^2} \mat N_t + \mat M_t \frac{\der^2 \HH}{\der \vct x^2} \mat M_t + \frac{\der^2 \HH}{\der \vct y\der \vct x} \mat N_t + \mat N_t \frac{\der^2 \HH}{\der \vct x\der \vct y}  - \frac{\der^2 \HH}{\der \vct y^2} \cr
\dot{\mM}_t & = &  - \mat M_t \frac{\der^2 \HH}{\der \vct x^2} \mat N_t  - \mat N_t \frac{\der^2 \HH}{\der \vct x^2} \mat M_t + \frac{\der^2 \HH}{\der \vct y\der \vct x} \mM_t  + \mM_t \frac{\der^2 \HH}{\der \vct x\der \vct y}  + \mat D \cr
\dot{a}_t & = & - \dot{\vct Y_t} \cdot \vct X_t -\HH \cr
\dot{b}_t & = & \frac{1}{2}\vct Y_t\cdot \mat D \vct Y_t,
\label{gausssystem}
\end{eqnarray}
where $\HH$ and all its derivative are implicitly taken at the point $\left(\vct X_t,\vct Y_t,t\right)$, and we used the notations $\mat D = \vct \lambda \vct \lambda^\top$ and
\begin{equation}
\frac{\der^2 \HH}{\der \vct x\der \vct y} = \left(
\begin{array}{cc}
\frac{\der^2\HH}{\der p\der y_p} & \frac{\der^2\HH}{\der p\der y_q} \cr
\frac{\der^2\HH}{\der q\der y_p} & \frac{\der^2\HH}{\der q\der y_q}
\end{array}
\right),
\end{equation}
and
\begin{equation}
\frac{\der^2 \HH}{\der \vct y\der \vct x} = \left(
\begin{array}{cc}
\frac{\der^2\HH}{\der p\der y_p} & \frac{\der^2\HH}{\der q\der y_p} \cr
\frac{\der^2\HH}{\der p\der y_q} & \frac{\der^2\HH}{\der q\der y_q}
\end{array}
\right) = \left( \frac{\der^2 \HH}{\der \vct x\der \vct y} \right)^\top,
\end{equation}
where $^\top$ means the transpose of a matrix or a vector.

We end up with a consistent system of ordinary differential equations. The first four equations are coupled, but the last two ones are actually trivial once the other are solved. 
Notice that these equations preserve the symmetry of $\mat M_t$ and $\mat N_t$ so we have ommited the transposition symbols that otherwise would appear in the expansion.

In the absence of environment, that is when $\mat D = 0$, the first two equations coincide with the double phase space trajectories of a classical dynamics, corresponding to a Liouville propagation of the chord function. The environment, expressed as an exponential damping centred on $\vct y = \vct 0$, then induces a shift of the maximum of the unitary evolved Gaussian which, otherwise, follows those trajectories. One should note that in the latter case, the semiclassical theory for Gaussian evolution becomes identical to the familiar unitary theory, albeit in an enlarged phase space. Thus, the pair of matrices that determine the squeezing and its direction interact because of the underlying classical motion. One can verify that equations (\ref{gausssystem}) are then consistent with the {\it Linearized Green's function and wavepacket propagation} of \cite{Heller}, as it is partially shown in appendix \ref{app_hell}. 

The qualitatively new feature of the present theory is that, unless $\mat D = 0$, the overall amplitude of the gaussian is damped by decoherence, if the Gaussian is not centred on the origin. It is visible from the formal expression of $\mat M_t$ solution of (\ref{gausssystem}):
\be
\mat M_t = \mat Q_t^\top\int_0^t \mat Q_{-\tau}^\top \mat D \mat Q_{-\tau}~d\tau ~\mat Q_t, 
\ee
where $\mat Q_t$ would be a time dependent $2\times 2$ matrix solution of
\be
\dot{\mat Q}_t = \mat Q_t\left(\frac{\der^2 \HH}{\der \vct y\der \vct x} - \frac{\der^2 \HH}{\der \vct x^2}\mat N_t\right).
\ee
This leads to
\be
\vct y\cdot\mat M_t\vct y= \int_0^t \left(\vct \lambda \cdot \mat Q_{t-\tau}\vct y\right)^2~d\tau, 
\ee
which is strictly growing. A Gaussian centred at the origin is merely squeezed by the Lindbladian term, leading to diffusive broadening of its Fourier transform, the Wigner  function, as described in \cite{BroAlm04, AlmBro07}.

It is also interesting to go back to see how these equations read in terms of the classical hamiltonian $H$:

\begin{eqnarray}
\dot{\vct X}_t^+ & = & \mat J \frac{\der H}{\der \vct x} \left(\vct X_t^+,t \right)  - \gamma\;\vct X_t^- + \left( \mat N_t + \frac{1}{2}\mat J \right) \mM_t^{-1}  \mat D \mat J \left( \vct X_t^+ - \vct X_t^- \right) \cr
\dot{\vct X}_t^- & = &  \mat J\frac{\der H}{\der \vct x} \left(\vct X_t^-,t \right)  - \gamma\;\vct X_t^+ + \left( \mat N_t - \frac{1}{2}\mat J \right) \mM_t^{-1}  \mat D \mat J \left( \vct X_t^+ - \vct X_t^- \right),
\label{gausssystem_H}
\end{eqnarray}
where $\vct X_t^\pm = \vct X_t \mp \frac{1}{2} \mat J \vct Y_t$. Once again, when there is no environment, i.e. $\mat D=\vct 0$ and $\gamma = 0$, the chord tips $\vct X_t^+$ and $\vct X_t^-$ are just independently following the time reverse classical motion
\begin{equation}
\dot{\vct X} = \mat J\frac{\der H}{\der \vct x}(\vct X,t).
\end{equation}

A general state can always be decomposed in a pseudo-basis of gaussian states:
\begin{equation}
\chi_0(\vxi) = \int D(\veta) 
\exp{\Bigl[
-\frac{(\vxi - \veta)\cdot M_{\veta}(\vxi - \veta)}{4\hbar} 
+ \frac{i}{\hbar}\phi_{\veta}
\Bigr]} 
~d\veta.
\label{generalinitial}
\end{equation}
Because of the linearity of the evolution equation of $\chi_t$, one can evolve each of these Gaussian states independently by using the consistent evolution presented in this section, and the superposition of these evolved Gaussians will be an approximate solution of (\ref{dynchi_der_H}) with the correct initial condition (\ref{generalinitial}). Although the expression of the latter is written as a continuous integral, one will obviously have to use a finite, approximate decomposition. About the issue of optimizing the number of Gaussian states of a decomposition, one may read with interest \cite{Ken04} and \cite{Maia07}.

\section{Quadratic Hamiltonian}

Here we discuss the consistent Gaussian evolution in the case of a quadratic Hamiltonian, namely
\begin{equation}
H(\vct x) = \vct x\cdot \mat H \vct x,
\label{quadham}
\end{equation}
with some symmetric matrix $\mat H$. The double phase space Hamiltonian is then
\begin{equation}
\HH(\vct x,\vct y) = -2\vct x\cdot\mat H \mat J \vct y - \gamma\; \vct x\cdot \vct y.
\end{equation}
It is obvious that the consistent evolution will give an exact result here, since the Hamiltonian coincides with its quadratic expansion. However it is instructive to explicit the calculus to have an insight on the role of the different terms.

As a reminder, the general solution derived in \cite{BroAlm04} was written in the following form 
\begin{equation}
\chi_t(\vct \xi) = \chi_0(e^{-\gamma t}\mat R_{-t}\vct \xi)\;\;  \exp{\left( 
-\frac{1}{2\hbar} \int_0^t e^{2\gamma(t'-t)} \left( \vct l 
\cdot \mat R_{t'-t}\vct \xi\right)^2  ~dt' \right)},
\label{exact-sol-chord}
\end{equation}
where $\chi_0(\vct \xi)$ is any general (not necessarily localized) initial function, and
\be
\mat R_t = \exp{\left(2\mat J \mat H t\right)}.
\ee
With the notation adopted in this paper, it reads
\begin{eqnarray}
A(\vct y,t) & = & A_0(\tilde{\vct y}_0) = 
A_0\left( e^{-\gamma t} \mat R_t^\top \vct y \right) \cr
B(\vct y,t) & = & B_0\left(e^{-\gamma t} \mat R_t^\top \vct y \right) + 
\frac{1}{2} \int_0^t{ e^{2\gamma(t'-t)} 
\left( \vct \lambda \cdot \mat R_{t-t'}^\top \vct y \right)^2} ~ dt'.
\end{eqnarray}
Here we used the correspondence
\be
\mat R_{-t} \vct \xi = - \mat R_{-t} \mat J \vct y = 
- \mat J \left( \mat R_t^\top \vct y \right).
\ee
If the initial state is a Gaussian wave packet, it imposes the form of $A_0(\vct y,t)$ and $B_0(\vct y,t)$, and one has
\begin{eqnarray}
A(\vct y,t)  & = &
a - (e^{-\gamma t} \mat R_t^\top \vct y-\vct Y)\cdot \vct X + \frac{1}{2} (e^{-\gamma t} \mat R_t^\top \vct y-\vct Y)\cdot\mat N (e^{-\gamma t} \mat R_t^\top \vct y-\vct Y)
\cr
B(\vct y,t) & = & b + \frac{1}{2} (e^{-\gamma t} \mat R_t^\top \vct y-\vct Y)\cdot\mM (e^{-\gamma t} \mat R_t^\top \vct y-\vct Y) + 
\frac{1}{2} \int_0^t{ e^{2\gamma(t'-t)} 
\left( \vct \lambda \cdot \mat R_{t-t'}^\top \vct y \right)^2} ~ dt'.
\label{ABquad}
\end{eqnarray}
This solution can be matched with (\ref{A_B_quad}) by setting the different parameters in the following way,
\begin{eqnarray}
\mat N_t & = & e^{-2\gamma t} \mat R_t \mat N \mat R_t^\top \cr
\mat M_t & = & e^{-2\gamma t} \mat R_t \mat M \mat R_t^\top + \int_0^t{ e^{2\gamma(t'-t)} \mat R_{t-t'} \mat D  \mat R_{t-t'}^\top }~ dt' \cr
\vct X_t & = & e^{-\gamma t} \mat R_t \left(\vct X + \mat N \vct Y\right) - \mat N_t  \vct Y_t \cr
\vct Y_t & = & \left(\mat M_t\right)^{-1} e^{-\gamma t} \mat R_{t} \mat M \vct Y \cr
a_t & = & a  + \vct X\cdot \vct Y + \frac{1}{2}\vct Y\cdot \mat N \vct Y - \vct X_t\cdot \vct Y_t - \frac{1}{2} \vct Y_t\cdot \mat N_t \vct Y_t \cr
b_t & = & b + \frac{1}{2} \vct Y \cdot \mat M\cdot \vct Y - \frac{1}{2} \vct Y_t \cdot \mat M_t \vct Y_t,
\label{coeffsquad}
\end{eqnarray}
where the matrix $\mat D$ is set for $\left( \vct \lambda \vct \lambda^\top \right)$; $\vct Y_t$ is defined as the maximum of the Gaussian, $\frac{\der B}{\der y}(\vct Y_t,t) = 0$, which is not the Hamiltonian evolution of the initial maximum $\vct Y$ of the Gaussian; and $\vct X_t$ is defined as $-\frac{\der A}{\der \vct y}(\vct Y_t,t)$, that is, the argument of the order $0$ of the expansion of $\vct \HH$ in (\ref{eq_ordre1}).

On the other hand we can explicit the consistent Gaussian evolution given by equations (\ref{gausssystem}): 
\begin{eqnarray}
\dot{\vct X}_t & = &  \left( 2 \mat J \mat H - \gamma\right) \; \vct X_t + \mat N_t (\mat M_t)^{-1} \mat D \vct Y_t \cr
\dot{\vct Y}_t & = & \left( 2 \mat H \mat J + \gamma\right) \; \vct Y_t  - (\mM_t)^{-1}  \mat D \vct Y_t \cr
\dot{\mat N}_t & = &  2  \mat J \mat H \mat N_t - 2 \mat N_t \mat H \mat J - 2 \gamma \mat N_t \cr
\dot{\mM}_t & = & 2  \mat J \mat H \mM_t - 2 \mM_t \mat H \mat J - 2 \gamma \mM_t  + \mat D \cr
\dot{a}_t & = &  \vct X_t \cdot (\mM_t)^{-1}  \mat D \vct Y_t  \cr
\dot{b}_t & = & \frac{1}{2}\vct Y_t\cdot \mat D \vct Y_t.
\label{gausssystemquad}
\end{eqnarray}
One can then check that (\ref{coeffsquad}) is solution of (\ref{gausssystemquad}).

The equations for the evolution of the coefficients of the Gaussian is far less intuitive than the simple interpretation of the solution (\ref{exact-sol-chord}). The latter is indeed a product of two Gaussians, each of which is fairly simple. The product of two Gaussians is itself a Gaussian, but its maximum does not obey any simple differential equation. Instead of (\ref{chigauss}), one could consider using an a priori solution structured as a product of two Gaussians, and perform the expansion of the Hamiltonian $\HH$ of (\ref{eq_ordre1}), not around $\vct Y = \vct Y_t$, but around both $\vct Y=\vct 0$ and $\vct Y = e^{-\gamma t}\mat R_t^\top \vct Y$. However, none of these points correspond to a zero of $\frac{\der B}{\der \vct y}(\vct Y,t)$ to justify the expansion around $\HH\left(-\frac{\der A}{\der \vct y}(\vct Y,t),t \right)$, which permits to separate the real from the imaginary part of (\ref{eq_ordre1}) in the Gaussian approach. This is not a problem in the quadratic case where the expansion is exact whatever the origin:
\begin{equation}
\frac{\der A}{\der t}(\vct y,t) +i \frac{\der B}{\der t}(\vct y,t) = 
 -\frac{\der A}{\der \vct y}(\vct y,t)
\cdot \left(2\mat H \mat J + \gamma\right) \vct y 
-  i\frac{\der B}{\der \vct y}(\vct y,t)
\cdot \left(2\mat H \mat J + \gamma\right) \vct y 
+ \frac{i}{2}( \vct \lambda\cdot\vct y )^2.
\label{eq_quad}
\end{equation}
In the non-quadratic case, this exact separation does not occur any more, and one must then relie on an awkward dynamics for $\vct Y_t$ and $\vct X_t$. However a classical dynamics can still be obtained by performing a complex WKB analysis, which will be addressed in \cite{BroAlm09}.

\section{Conclusion}

We have derived an approximate Gaussian solution of the Lindblad equation, based on a second order expansion of the Hamilton Jacobi equation of the phase of the chord function. The trajectory of the maximum of the Gaussian does not follow an intuitive trajectory, as it is a compromise between the unitary evolution of the Gaussian, and the damping induced by the Lindbladian term. The latter quickly shrinks the relevant part of the chord function to a region close to $\vct y = \vct 0$ on the $\sqrt{\hbar}$ scale, which can be interpreted as the region containing all the classical information about the state. Thus the maximum of the Gaussian chord function ends up being at $\vct y =\vct 0$. This result is in agreement with established results in the case of a quadratic Hamiltonian and linear Lindblad operators, see for instance \cite{BroAlm04}, as well as fulfilling our handwaving intuition that ``decoherence drives the system back into classical dynamics''. However, with a nonquadratic Hamiltonian, the rate of this environmentally induced exponential damping acquires a phase space dependence.

Our theory generalizes the standard semiclassical approximations for the unitary evolution of wave packets \cite{Heller} in a nontrivial way: in the chord representation adopted here, a single wave packet is always centred on the origin (its position is represented by an overall phase). Thus, by allowing gaussians away from the origin, like in the typical case of a ``Shr\"odinger cat'' state recalled in appendix \ref{app_cat}, the approximate gaussian evolution described by the theory includes the interferences between pairs of wave packets, their motion and their loss of coherence.

\appendix

\section{Asymptotic expansion of the Hamiltonian operator}
\label{appa}

One can verify easily that
\begin{eqnarray}
\left( \frac{1}{\lambda} \frac{\der}{\der p} \right)^m
\left( \frac{1}{\lambda} \frac{\der}{\der q} \right)^n \exp{\lambda S} & = &
\Biggl[
\left( \frac{\der S}{\der p} \right)^m \left( \frac{\der S}{\der q} \right)^n
+ \frac{1}{\lambda} 
\left( 
\frac{m(m-1)}{2} \frac{\der^2 S}{\der p^2} \left( \frac{\der S}{\der p} \right)^{m-2}
\left( \frac{\der S}{\der q} \right)^n  
\right. 
 \cr 
~ & ~ & 
\left.
+ mn \frac{\der^2 S}{\der p \der q} \left( \frac{\der S}{\der p} \right)^{m-1} \left( \frac{\der S}{\der q} \right)^{n-1} + \frac{n(n-1)}{2} \frac{\der^2 S}{\der q^2} \left( \frac{\der S}{\der p} \right)^{m}
\left( \frac{\der S}{\der q} \right)^{n-2}
\right) 
 \cr
~ & ~ & 
+ \GO(\frac{1}{\lambda^2})
\Biggr] \exp{\lambda S}.
\end{eqnarray}
From there one can generalize and write
\begin{equation}
\HH\left( -\frac{\hbar}{i}\frac{\der}{\der \vct y}, \vct y \right)
\exp{\left(\frac{i}{\hbar}S \right)} = \Biggl[
\HH\left(-\frac{\der S}{\der \vct y}, \vct y \right) + \frac{\hbar}{2i}\mathop{Tr}{\{\frac{\der^2 \HH}{\der \vct x^2}\left(-\frac{\der S}{\der \vct y}, \vct y \right)
\frac{\der^2 S}{\der \vct y^2}\}} + \GO(\hbar^2)
\Biggr]\exp{\left(\frac{i}{\hbar}S \right)}
\end{equation}

\section{Consistence with Heller's equations}
\label{app_hell}

In the paragraph {\it 2.3. Linearized Green's function and wavepacket propagation} of \cite{Heller}, Heller starts with a Gaussian wave function with
\be
\psi_h(q,t) = \exp{\{i/\hbar[A(q-Q)^2 + P(q-Q) + s_t]\}},
\label{stateheller}
\ee
where $A = a+ib$ is a time dependent complex number, P a real time dependent number and we have sticked to the $1$ dimensionnal case. Then he proceeds with
\bea
\dot{Q} & = & \frac{\der H}{\der p} \cr
\dot{P} & = & - \frac{\der H}{\der q} \cr
A & = & \frac{p_z}{2z} \cr
\dot{p_z} & = & - V'' z \cr
\dot{z} & = &  \frac{1}{m}p_z,
\label{evolheller}
\eea
where $H$ stands for $H(P,Q) = \frac{P^2}{2m} + V(Q)$, and $V''$ for $\frac{d^2 V}{dQ^2}(Q)$. This defines the consistent Gaussian dynamics of (\ref{stateheller}) according to Heller.

In this appendix we show that the evolution of the chord representation $\chi_{h}(t)$ of the state $|\psi_h(t)\rangle\langle \psi_h(t)|$, as defined above, also follows equations (\ref{gausssystem}) in the unitary case $\mat D = \vct 0$. This shows in a simple case that our consistent Gaussian evolution is consistent with the one of Heller in the unitary case. Both theories also agree for a general state, although the derivation, carrying much heavier expressions, is not written out in this appendix. 

From (\ref{stateheller}) and (\ref{chordfunc}) one has
\be
\chi_{h}(t) = \exp{\frac{i}{\hbar}\left(
-P y_p - Q y_q + s_t - s_t^*
+\frac{i}{2} \vct y\cdot \mat M \vct y
\right)},
\ee
with
\be
\mat M = \left(\begin{array}{cc} \displaystyle
b+\frac{a^2}{b} & \displaystyle \frac{a}{2b} \cr
\displaystyle \frac{a}{2b} 
& \displaystyle \frac{1}{4 b} \end{array}\right).
\ee
Notice already that the first two lines of (\ref{evolheller}) give the second line of (\ref{gausssystem}) with $\vct X_t = (P,Q)$. On the other hand $\vct Y_t = \vct 0$ and $\mat N_t = \vct 0$ in this case.
Moreover, the equation for $A$ in (\ref{evolheller}) gives
\be
\dot{\mat M} = \left(\begin{array}{cc} \displaystyle
-\frac{4}{m}ab-\frac{2a}{b}\left(\frac{V''}{2}+\frac{2}{m}(a^2-b^2)\right) + \left(\frac{a}{b}\right)^2\frac{4}{m}ab
& \displaystyle -\frac{1}{2b} \left(\frac{V''}{2}+\frac{2}{m}(a^2-b^2)\right) + \frac{a}{2b^2}\frac{4}{m}ab
\cr
\displaystyle  -\frac{1}{2b} \left(\frac{V''}{2}+\frac{2}{m}(a^2-b^2)\right) + \frac{a}{2b^2}\frac{4}{m}ab
& \displaystyle \frac{1}{4 b^2}\frac{4}{m}ab \end{array}\right),
\ee
that is
\be
\dot{\mat M} = \left(\begin{array}{cc}
-2V'' \mat M_{qq} & -V'' \mat M_{qq} + \frac{1}{m}\mat M_{pp} \cr
-V'' \mat M_{qq} + \frac{1}{m}\mat M_{pp} & \frac{2}{m}\mat M_{pq}
\end{array}\right).
\ee
It is equivalent to the fourth line of (\ref{gausssystem}), with
\be
\frac{\der^2 \HH}{\der \vct y\der \vct x} = \left(
\begin{array}{cc}
0 & - V''(Q) \cr
\frac{1}{m} & 0
\end{array}
\right). 
\ee
We do not take into account $s_t$ which corresponds to the prefactor of $\chi$, which would be given by the next order in $\hbar$ in our expansion.

To check the general case, with $\vct Y_t \neq \vct 0$ and $\mat N_t \neq \vct 0$, one must write the chord representation $\chi^h_{ab}$ of the cross product $|\psi^h_a(t)\rangle\langle \psi^h_b(t)|$ of two different Gaussian states which follow equations (\ref{evolheller}). The expression of the time derivative of that chord function can be shown to be consistent with (\ref{gausssystem}) in the same way as what is done above, but quite heavier.

\section{``Schr\"odinger cat'' states}
\label{app_cat}

The generalized ``cat state'', that is a coherent superposition of two squeezed states, is an important example for two reasons. First it is Gaussian, and therefore is adapted to the treatment of this article; and secondly it gives a quite transparent physical interpretation of the chord function. We define the two ingredients of such a state: two squeezed states
\begin{eqnarray}
\psi_a(q) & = & \frac{1}{(\pi\hbar\omega_a^2)^{1/4}}
\exp{\left(
-\frac{(q-Q_a)^2}{2\omega_a^2\hbar}+\frac{iP_a q}{\hbar}
\right)} \cr
\psi_b(q) & = & \frac{1}{(\pi\hbar\omega_b^2)^{1/4}}
\exp{\left(
-\frac{(q-Q_b)^2}{2\omega_b^2\hbar}+\frac{iP_b q}{\hbar}
\right)},
\end{eqnarray}
centred respectively on $\vct X_a = (P_a,Q_a)$ and $\vct X_b = (P_b,Q_b)$.
The density operator of the corresponding ``cat state'', $|\psi_a\rangle + |\psi_b\rangle$, has two diagonal terms and two off-diagonal terms:
\begin{equation}
\op \rho = \frac{1}{2}|\psi_a\rangle \langle\psi_a| + 
\frac{1}{2}|\psi_b\rangle \langle\psi_b| + 
\frac{1}{2}|\psi_a\rangle \langle\psi_b| 
+ \frac{1}{2}|\psi_b\rangle \langle\psi_a|
\end{equation}
We are interested here in the sum of the pair of non-local terms, $|\psi_a\rangle \langle\psi_b|$ and complex conjugate. Notice that the two other diagonal terms, which have a classical interpretation, can be retrieved anyway by setting $a=b$. In the Weyl representation, this non-local term can be written
\begin{eqnarray}
W_{ab}(\vct x) & = &
\frac{1}{\pi\hbar}\sqrt{\frac{2\omega_a\omega_b}{\omega_a^2+\omega_b^2}}
\exp{
\left(
-\frac{(q-Q)^2 + \omega_a^2\omega_b^2(p-P)^2 
+ i(\omega_b^2-\omega_a^2)(p-P)(q-Q)}
{2\hbar(\omega_a^2+\omega_b^2)} \right. } \cr
~ & ~ & 
{ \left. + \frac{i}{\hbar}\left( - \vct Y\cdot \vct x - P Y_p \right)
\right)},
\label{wignercat}
\end{eqnarray}
where $\vct X = (P,Q) = \frac{\vct X_a + \vct X_b}{2}$ and $\mat J \vct Y = \vct X_b - \vct X_a$.
In the chord representation, one has
\begin{eqnarray}
\chi_{ab}(\vct y) & = &
\frac{1}{2\pi\hbar}\sqrt{\frac{2\omega_a\omega_b}{\omega_a^2+\omega_b^2}}
\exp{
\left(
-\frac{(y_p-Y_p)^2 + \omega_a^2\omega_b^2(y_q-Y_q)^2 
+ i(\omega_b^2-\omega_a^2)(y_p-Y_p)(y_q-Y_q)}
{2\hbar(\omega_a^2+\omega_b^2)} \right. } \cr
~ & ~ & 
{ \left. + \frac{i}{\hbar}
\left(-\vct X\cdot \vct y + Y_q Q \right)
\right)}.
\end{eqnarray}
Then one has
\begin{equation}
A_0(\vct y) = \frac{1}{2}\frac{\omega_a^2-\omega_b^2}
{\omega_a^2+\omega_b^2}(y_p-Y_p)(y_q-Y_q)
-\vct X\cdot \vct y + Y_q Q,
\label{A0cat}
\end{equation}
and
\begin{equation}
B_0(\vct y) = \frac{1}{2}\frac{(y_p-Y_p)^2 + \omega_a^2\omega_b^2(y_q-Y_q)^2 }
{\omega_a^2+\omega_b^2}.
\label{B0cat}
\end{equation}
One may have noticed that $A_0(\vct y)$ is not an odd function, nor $B_0(\vct y)$ an even function, but one has to bear in mind that we looked only at $\chi_{ab}(\vct y)$, and that the complete chord function contains also $\chi_{ba}(\vct y)$, which re-establishes the overall symmetry, $\chi(\vct y) = \chi(-\vct y)^*$ for Hermitian operators.

\begin{acknowledgments}

We thank Raul Vallejos and Denis Ullmo for helpful comments on the manuscript.
Partial financial support from Millenium Institute of Quantum
Information, FAPERJ, PROSUL, CNPq and CAPES-COFECUB is gratefully acknowledged.

\end{acknowledgments}

\end{document}